\documentclass[conference]{IEEEtran}
\ifCLASSINFOpdf
\else
\fi

\usepackage{float}
\usepackage{graphicx}
\usepackage{color}
\usepackage[square, comma, sort&compress, numbers]{natbib}
\usepackage{subfigure}
\usepackage{amsmath}
\usepackage{algorithm}
\usepackage{algorithmic}
\usepackage{amssymb}
\usepackage{threeparttable}
\usepackage{booktabs}
\usepackage{makecell}
\usepackage{colortbl}
\usepackage{amsfonts}
\usepackage{amssymb}
\usepackage{hyperref}
\usepackage{url}
\usepackage{mathtools}
\usepackage{cuted}
\usepackage{multicol}
\usepackage{stfloats}
\usepackage{array,multirow}

\usepackage{graphicx}
\definecolor{mygray}{gray}{.9}
\begin{document}
\title{Auto-Generation of Pipelined Hardware Designs for Polar Encoder}
\author{\IEEEauthorblockN{Zhiwei Zhong$^{1,2}$, Xiaohu You$^{2}$, and Chuan Zhang$^{1,2,*}$}
\IEEEauthorblockA{$^{1}$Lab of Efficient Architectures for Digital-communication and Signal-processing (LEADS)\\
$^{2}$National Mobile Communications Research Laboratory, Southeast University, Nanjing, China\\
Email: \{zwzhong, xhyu, chzhang\}@seu.edu.cn
}

}

%

\maketitle
\begin{abstract}
This paper presents a general framework for auto-generation of pipelined polar encoder architectures. The proposed framework could be well represented by a general formula. Given arbitrary code length $N$ and the level of parallelism $M$, the formula could specify the corresponding hardware architecture. We have written a compiler which could read the formula and then automatically generate its register-transfer level (RTL) description suitable for FPGA or ASIC implementation. With this hardware generation system, one could explore the design space and make a trade-off between cost and performance. Our experimental results have demonstrated the efficiency of this auto-generator for polar encoder architectures.
\end{abstract}
\IEEEpeerreviewmaketitle
\section{Introduction}
Polar code \cite{arikan2009channel}, the first channel code which can provably achieve the capacity of the binary-input discrete memoryless channels (BDMCs), has been considered as the recent breakthrough of coding theory. Recently, polar code has been adopted by the enhanced mobile broadband (eMBB) control channels for the $5$G NR interface. As pointed out by \cite{arikan2009channel}, to achieve a good error-correcting performance of polar code, the code length is expected to be sufficiently long. However, as for polar code, the hardware complexity of fully parallel encoder will be high as the code length increases. Therefore, pipelined architecture should be introduced to reduce the hardware cost. Using folding transformation \cite{parhi1992synthesis}, \cite{zhang2015pipelined} has proposed both feed-forward and feed-back polar encoder with $2$-parallel processing; \cite{yoo2015partially} has proposed pipelined polar encoder architecture with $4$-parallel processing. Although \cite{yoo2015partially} has claimed that the folding transformation could derive polar encoder with any level of parallelism, the detailed framework is not given.

In synthesizing hardware architectures for an $N$-bit polar encoder, different level of parallelism leads to different latency, throughput, silicon area and memory cost. Intuitively, the level of parallelism $M$ suitable for an $N$-bit polar encoder should be $2 \leqslant M \leqslant N/2$, where $M$ is a power of two. Thus, as the code length increases, there will be more choices of $M$ and the design space will be wider. Therefore, it will be exhausting to choose the optimal values of $N$ and $M$ under different hardware constraints.

In order to fulfill the requirements of different applications, a auto-generator which can connivently output polar encoder architecture with given code length $N$ and parallelism $M$ is highly expected. Also, this auto-generator can free the hardware designers from the laborious case designs, bypass the hardware details, and give the design space in a more convenient way. Inspired by a fast Fourier transform (FFT) generator  \cite{milder2012computer} which could automatically generate FFT hardware architecture with arbitrary parallelism and figure out hardware cost, this paper proposes an auto-generation system which could produce polar encoder hardware architecture with arbitrary code length and arbitrary level of parallelism.

The remainder of this paper is organized as follows. In Section \ref{sec:1}, the brief description of polar encoding is introduced. In Section \ref{sec:2}, we propose the generation system of polar encoder and an exemplary $32$-bit polar encoder with $8$-parallel processing. In Section \ref{sec:3}, the analysis of the performance of the generation system is given. In Section \ref{sec:4}, we conclude and remark on the entire paper.

\section{Preliminaries}\label{sec:1}
\subsection{Polar Encoder}
In polar code encoding, $u_{0}^{N-1}$ is regarded as the source word and $x_{0}^{N-1}$ as the codeword. The encoding scheme can be defined by Eq. $(1)$, where $G_{N}$ and $B_{N}$ are the generation matrix and the bit-reversal permutation matrix respectively, and $F^{\otimes n}$ is the \emph{Kronecker power} of \(n\) with $n=\log_{2}N $ and $F = \left[\begin{smallmatrix} 1 & 0 \\ 1 & 1 \end{smallmatrix}\right]$.
\begin{equation}\label{eq:n1}
    \begin{array}{ll}
      x_{0}^{N-1} = u_{0}^{N-1} G_{N}= u_{0}^{N-1} B_{N} F^{\otimes n}.
	\end{array}
\tag{1}
\end{equation}

\begin{figure}[ht]
\centering
\includegraphics[width=.8\linewidth]{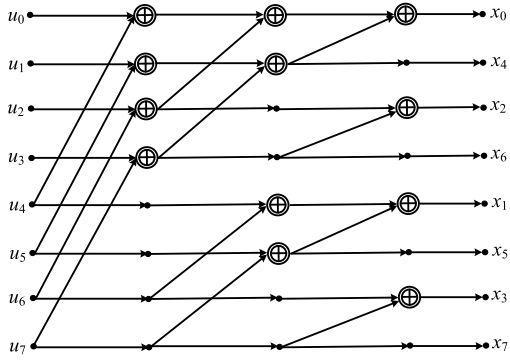}
\caption{The data-flow graph of an $8$-bit polar encoder.}\label{fig:1}
\end{figure}

As proved by \cite{zhang2015pipelined}, the data-flow graph (DGF) of polar encoder could be derived from the DFG of FFT processors by replacing all the butterfly modules with \textsc{xor}-and-\textsc{pass} modules, and all the twiddle factors with 1's. Therefore, the proposed framework for polar encoder has the potential for implementing the pipelined hardware architecture for FFT by reversing the replacement. An exemplary DGF of an $8$-bit polar encoder is shown in Fig. \ref{fig:1}. Note that this DFG is similar to the that of an $8$-point radix-$2$ decimation-in-frequency (DIF) FFT processor in the way mentioned above.

\section{Hardware Generation}\label{sec:2}
\begin{figure}[ht]
\centering
\includegraphics[width=.8\linewidth]{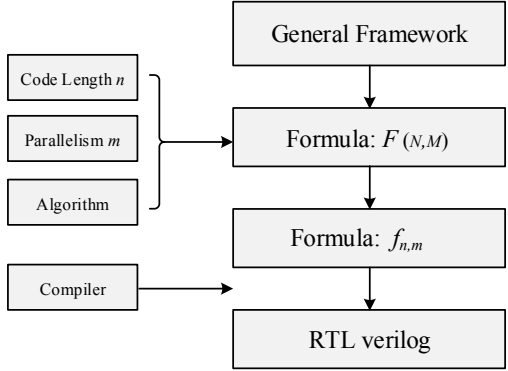}
\caption{The hardware generation system for polar encoder.}\label{fig:4}
\end{figure}

In this section, we introduce the general pipelined framework for polar encoder with arbitrary code length $N$ and arbitrary level of parallelism $M$. The general framework could be easily denoted by a general formula \(F(N, M)\). Then we show how to use an algorithm to derive a specific formula \(f_{N, M}\) from \(F(N, M)\) based on the values of $N$ and $M$. Finally, a compiler is employed to translate \(f_{N, M}\) into RTL description. The hardware generation system is illustrated in Fig. \ref{fig:4}.

\subsection{From General Framework to Formula}
\begin{figure}[ht]
\centering
\includegraphics[width=\linewidth]{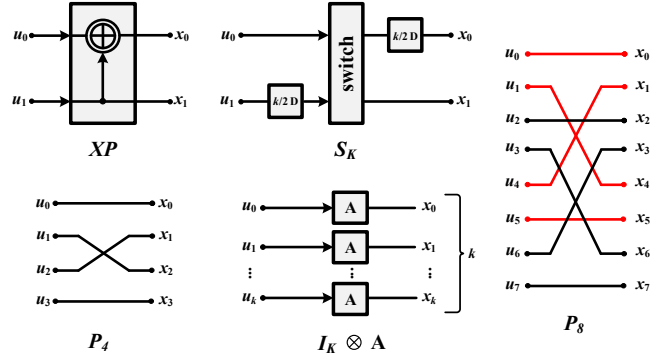}
\caption{The symbols and corresponding hardware modules in the formula.}\label{fig:2}
\end{figure}

Consider that the general framework is expected to implement polar encoder with arbitrary code length and arbitrary level of parallelism, the framework should be scalable, i.e., the number of stages and the number of hardware modules in each stage should change with the values of $N$ and $M$.
Such a scalable framework could represented by formula \(F(N,M)\) shown in Eq. $(2)$. Here the parameters $N$ and $M$ are powers of $2$, and $4 \leqslant M \leqslant N/2$. Before we go into details of \(F(N,M)\), we introduce all the symbols that might be used in \(F(N,M)\) and \(f_{N,M}\), as well as the symbols' corresponding hardware modules. Note that the final hardware implementation of \(f_{N,M}\) is the serial connection of the individual modules of different symbols. Fig. \ref{fig:2} illustrates all the exemplary modules, as well as symbols, that might be used in our design, all of which take \(u_{0}^{N-1}\) as input and \(x_{0}^{N-1}\) as output.

Symbol $\textit{XP}$ represents an \textsc{xor}-and-\textsc{pass} module that achieves: $x_0 = u_0 + u_1$ (in GF($2$)) and $x_1 = u_1.$ The number of inputs of $\textit{XP}$ is fixed and equals to $2$ in our design.

Symbol $S_K$ ($K$ is a power of $2$, $K>1$) represents a switch with $k/2$ delay elements (denoted by $D$) on each side. A $\log_{2}K$-bit counter is needed to control the switch: the value $0$ of the most significant bit of the counter infers direct data transfer, and the value $1$ infers cross data transfer. The number of inputs of $S_K$ is fixed and equals to $2$.

Symbol $P_N$ ($N$ is a power of $2$, $N>2$) denotes the permutation on an $N$-dimensional vector. The detail function of $P_N$ is illustrated in Algorithm \ref{alg1}. Intuitively, $P_N$ is the duplication of $P_{N/2}$. For example, $P_8$ could be viewed as partial overlap of two $P_4$ modules with red wires and black wires respectively.

Symbol \textit{$({I_K}\otimes{A}$)} ($K$ is a power of $2$, $K>0$) is a Kronecker product representing $K$ parallel instances of module $A$, where $A$ is an abstract module and $A$ could be replaced by $\textit{XP}$, $S_K$ or $P_N$. Note that when $K=1$, \textit{$({I_K}\otimes{A})$} equals to ${A}$. Suppose that $A$ has $X$ inputs, the number of inputs of \textit{$({I_K}\otimes{A}$)} equals $K \times X$.

The general formula $F(N,M)$ is composed of symbols mentioned above, except that the $W$ in Eq. $(2)$ is a variable module. When deriving $f_{N,M}$ from $F(N,M)$, symbol $W$ should be replaced by $P_N$ or $S_K$ according to its subscript. In Algorithm \ref{alg2}, as the code length and the level of parallelism are given, all the subscripts of each symbol in $F(N,M)$ will be figure out. Then the module {\small\text{$({I}\otimes{W})$}} is replaced by {\small\text{$({I}\otimes{P})$}} or {\small\text{$({I}\otimes{S})$}} based on the value of the subscript of $W$. Finally, the formula $f_{N,M}$ is determined.

\subsection{Compiler}
We have built a compiler in \textit{Python} that takes $f_{N,M}$ as input and automatically connects all the basic modules in $f_{N,M}$ in left-to-right order. Specifically, as we input $N$ and $M$ into $F(N,M)$, the $f_{N,M}$ will be determined and transformed into the register-transfer level (RTL) Verilog by the compiler. The detail of the compiler is beyond the scope of this paper; we only provide a brief introduction here.

There are totally three types of basic modules in the formula $f_{N,M}$: the \textsc{xor}-and-\textsc{pass} module \textit{XP}, the switch module $S_K$, and the permutation module $P_N$. There are two ways to expand these modules. The first one is to employ the symbol ${I_{K}\otimes}$ to layout the duplication of one module in a parallel way. The other one is to change the symbols' subscripts. Therefore, the compiler needs to read each symbol of $f_{N,M}$ from left to right, and recognizes ${I_{K}\otimes}$ as well as each symbol's subscript. Then the compiler could determine the specific hardware architecture and print the Verilog files.

\begin{figure*}[htbp]
\hrulefill
\begin{center}
\begin{align*}
({I_{M/2}}\otimes{\textit{XP}})({I_{M/4}\otimes{P_4}}) \left\lbrace \Pi_{i=0}^{\log_{2N-3}}[({I_{M/2}}\otimes{W_{N/(2^{i}M)}})({I_{M/2}}\otimes{\textit{XP}})] \right\rbrace ({I_{M/4}\otimes{P_4}}) ({I_{M/2}} \otimes {S_{N/M}})({I_{M/2}}\otimes{\textit{XP}})   \qquad  (2)
\end{align*}
\begin{align*}
({I_4}\otimes{\textit{XP}})({I_2\otimes{P_4}}) \left\lbrace ({I_{4}}\otimes{W_4})({I_{4}}\otimes{\textit{XP}})({I_{4}}\otimes{W_2})({I_{4}}\otimes{\textit{XP}}){W_1} ({I_{4}}\otimes{\textit{XP}}) \right\rbrace ({I_{2}\otimes{P_4}}) ({I_{4}} \otimes {S_{4}})({I_{4}}\otimes{\textit{XP}}) \qquad \qquad  \ \ \ \ (5)
\end{align*}
\begin{align*}
({I_4}\otimes{\textit{XP}})({I_2\otimes{P_4}}) \left\lbrace ({I_{4}}\otimes{S_4})({I_{4}}\otimes{\textit{XP}})({I_{4}}\otimes{S_2})({I_{4}}\otimes{\textit{XP}}){P_8} ({I_{4}}\otimes{\textit{XP}}) \right\rbrace ({I_{2}\otimes{P_4}}) ({I_{4}} \otimes {S_{4}})({I_{4}}\otimes{\textit{XP}}) \qquad \qquad \qquad \,  \ (6)
\end{align*}
\end{center}
\end{figure*}

\subsection{Input and Output Orders}
The input and output data of this framework are in regular order. Suppose the input data of $f_{N,M}$ is $u_0^{N-1}$, since $f_{N,M}$ represents a pipelined architecture, $u_0^{N-1}$ will be divided into $N/M$ $M$-dimensional vectors $V_{in(i)}$ illustrated in Eq. $(3)$, where $i=0,1,...,{(N/M)-1}$. All the data in $V_{in(i)}$ will be entered into the encoder in parallel, and $i$ indicates the sequence of the input vector, i.e., $V_{in(0)}$ is the first set of the input data and the $V_{in(N/M-1)}$ is the last set of the input data. The output data are in bit-reversal order. Specifically, suppose $x_0^{N-1}$ is the theoretical codeword and $y_0^{N-1}$ is in the bit-reversal form of $x_0^{N-1}$. Then the $i$-th output vector $V_{out(i)}$ equals to $y_{M \times i}^{(M \times i)+(M-1)}$, where $i=0,1,...,(N/M)-1$.

\begin{equation}
\mathbf{V}_{in(i)}=
\resizebox{.4\hsize}{!}{$
\begin{bmatrix}
  u_{(M/2) \times i} \\
  u_{(M/2) \times i+(N/2)} \\
  u_{(M/2) \times i+1} \\
  u_{(M/2 \times i)+1+(N/2)} \\
  u_{(M/2) \times i+2} \\
  u_{(M/2 \times i)+2+(N/2)} \\
  ... \\
  u_{(M/2) \times i+(M/2)-1} \\
  u_{(M/2) \times i+(M/2)-1+(N/2)}
\end{bmatrix}$}.
  \tag{3}
\end{equation}

For the general framework, the processing latency (clock cycles) is {\small\text{$T_{latency} = (3N/2M) - 1$}}. The number of \textsc{xor} gates and delay elements are:
\begin{equation}
    \begin{array}{ll}
    ^\# \textit{XOR} = (M/2) \times  \log _{2}N;\\
     ^\# \textit{MEM} = (3N/2) - M.
	\end{array}
\tag{4}
\end{equation}

\begin{algorithm}
\caption{The Permutation on an $N$-dimensional Vector}
\label{alg1}
\begin{algorithmic}[1]
\REQUIRE The input vector $u_{0}^{N-1}$.
\FOR{$(i=0;i<N/2;i=i+2)$}
\STATE $x[i] = u[i]$
\ENDFOR
\;
\FOR{$(i=N-1;i>N/2;i=i-2)$}
\STATE $x[i] = u[i]$
\ENDFOR
\;
\FOR{$(i=1;i<N/2;i=i+2)$}
\STATE $x[i] = u[i-1+(N/2)]$
\STATE $x[i-1+(N/2)] = u[i]$
\ENDFOR
\;
\STATE Output $x_{0}^{N-1}$.
\end{algorithmic}
\end{algorithm}

\begin{algorithm}
\caption{The Generation of Formula $f_{N,M}$}
\label{alg2}
\begin{algorithmic}[1]
\REQUIRE The code length $N$ and the level of Parallelism $M$ ($N = 2^i, M = 2^i, i\geqslant 2, i\in Z, M \leqslant N/2$).
\STATE Input $N$ and $M$ in to the general formula \(F(N,M)\).
\;
\IF{$(k>=1,k=2^i,i\in Z)$}
\STATE $({I_{M/2}} \otimes W_{1/k})=(I_{k}\otimes P_{M/k})$

\ELSE
\STATE $({I_{M/2}} \otimes W_{1/k})=({I_{M/2}} \otimes S_{1/k})$
\ENDIF
\;
\STATE $f_{N, M} = F(N,M)$
\;
\STATE Output the formula \(f_{N, M}\).
\end{algorithmic}
\end{algorithm}

\subsection{A 32-Bit 8-Parallel Polar Encoder}
\begin{figure*}[t]
\centering
\includegraphics[width=\linewidth]{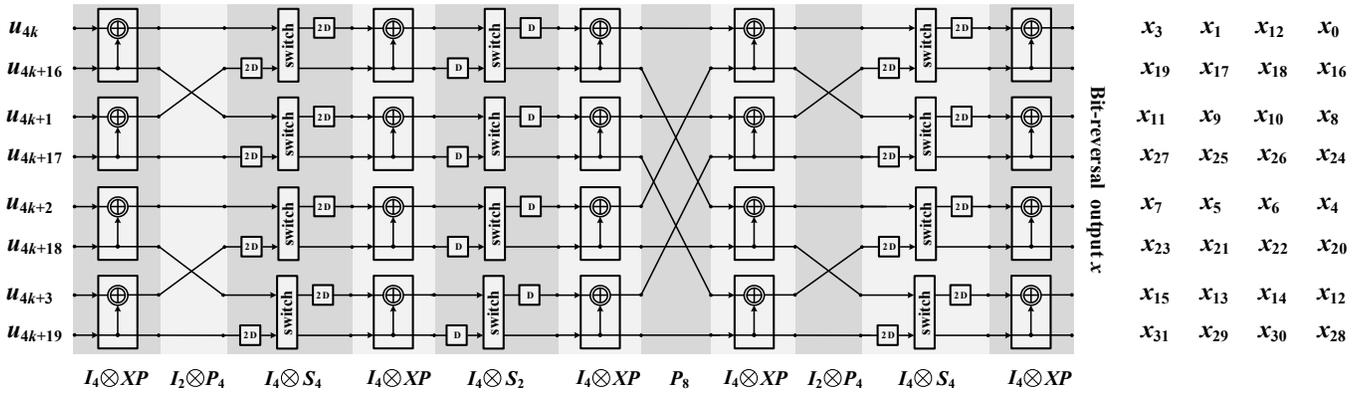}
\caption{The hardware architecture of polar encoder with $N=32$, $M=8$.}\label{fig:3}
\end{figure*}

According to Algorithm \ref{alg2}, given $N=32$ and $M=8$, formulas $F(32,8)$ and $f_{32,8}$ are obtained in Eq. $(5)$ and Eq. $(6)$, respectively. The hardware architecture is illustrated in Fig. \ref{fig:3}, which consists of $20$ \textsc{xor} gates and $40$ delay elements in accordance with Eq. $(4)$. The architecture could be split in $11$ columns, each of which has its relevant symbol under the column. Note that Eq. $(6)$ is actually composed of all the symbols at the bottom of Fig. \ref{fig:3}. The order of the input data $u$ ($k=0,1,2,3$) at the leftmost part of Fig.  \ref{fig:3} conforms to the order mentioned above. The output data $x$ is in the bit-reversal order.

\section{Performance and Complexity}\label{sec:3}

Some of the hardware designs derived from the auto-generation system were implemented on the Xilinx Virtex-7
VC709 FPGA platform with Virtex-7 XC7VX690T. All the design examples are of the same code length $N=1024$, but with different level of parallelism. The synthesis results are illustrated in Table \ref{table1}. From the table, it can be observed that the throughput (T/P) and the number of Slice LUTs and Slice Registers increase as the value of $M$ increases. In an extreme case, the polar encoder with $M=512$ consumes more Slice LUTs than the polar encoder with $M=4$ by $5167\%$ but achieves higher throughput by $8710\%$.

As mentioned in Section \ref{sec:2}, the value of $M$ conforms to $4 \leqslant M \leqslant N/2$. Then, given the code length $N$, the generation system could implement $(\log_{2}N)- 2$ designs with different $M$, covering a wide cost/performance trade-off space. Therefore, one could choose the most suitable polar encoder in the design space to fit the application.
\begin{table}[ht]
\caption{Implementation of The Hardware Designs Derived From The Auto-Generation System on the Xilinx Virtex-7 VC709 FPGA Platform with Virtex-7 XC7VX690T.}
\centering
\begin{tabular}{c c c c c c}
\Xhline{0.8pt}

\multirow{2}{*}{$N$} &\multirow{2}{*}{$M$} & \multirow{2}{*}{Slice LUTs} & \multirow{2}{*}{Slice Registers} & \text{Max freq} & \text{T/P} \\
\textbf{} & \textbf{} & \textbf{} & \text{} & \text{(MHz)} & \text{(Gbps)} \\
\hline
\textbf{$1024$} & \textbf{$4$} & \textbf{$148$} & \textbf{$82$}  & \textbf{$519.535$} & \textbf{$2.07
$}\\

\textbf{$1024$}& \textbf{$32$} & \textbf{$467$}  & \textbf{$312$}& \textbf{$407.05$} & \textbf{$13.02$}\\
\textbf{$1024$} & \textbf{$128$} & \textbf{$1278$} & \textbf{$845
$}& \textbf{$340.518$} & \textbf{$43.58$}\\
\textbf{$1024$} & \textbf{$256$} & \textbf{$1704$} & \textbf{$1194
$}& \textbf{$348.712$} & \textbf{$89.27$}\\
\textbf{$1024$} & \textbf{$512$} & \textbf{$2628$}& \textbf{$1025
$} & \textbf{$356.223$} & \textbf{$182.38$}\\
\Xhline{0.8pt}
\end{tabular}
\label{table1}
\end{table}

\section{Conclusion}\label{sec:4}
This paper proposes an auto-generation system for the hardware architecture of polar encoder. The system could offer users a wide range of design space so that the users could make a trade-off between cost and performance to best fit their applications. The essence of the generation system lies in the formula-based expression of the general framework for polar encoder that could achieve encoding with arbitrary code length and arbitrary parallelism. This auto-generation can help designers to conveniently design polar encoder without touching hardware details. The derivation of design space can further help us to identify the required design.

In this paper, we also introduce the scalable hardware modules associated with the formula, as well as the compiler that could transform the formula into RTL Verilog files. Synthesis results on FPGA have demonstrated the efficiency and the large trade-off space of the auto-generated circuits.

Future work will be directed toward the auto-generation of successive cancellation polar decoder and belief prorogation decoder based on our previous works \cite{zhang2012reduced,zhang2013low,yang2016bp}, and the design optimization based on the design space.



\small

\bibliographystyle{IEEEtran}
\bibliography{IEEEabrv,mybib}
\end{document}